\documentclass[a4paper, 12pt]{report}
\usepackage{amsfonts,,amsthm, amssymb,array,hyperref, amsmath}
\newcolumntype{C}[1]{>{\centering\let\newline\\\arraybackslash\hspace{0pt}}m{#1}}
\newcolumntype{R}[1]{>{\flushright \let\newline\\\arraybackslash\hspace{0pt}}m{#1}}
\usepackage[hmargin=3.5cm,vmargin=3cm]{geometry}
\title{Deterministic Primality Testing}
\author{Vijay Menon}
\date{}
\begin{document}
\maketitle
\pagenumbering{roman}
% \begin{center}
% \textbf{{\large ACKNOWLEDGEMENT}}
% \end{center}
% 
% \addcontentsline{toc}{chapter}{Acknowledgement}
% 
% \noindent \normalsize \\ Foremost, I would like to express my sincere gratitude to my guide Dr. Ayineedi Venkateswarlu for giving me an opportunity to work under him, for his motivation, encouragement, and continuous support through out this internship. His valuable guidance and immense knowledge helped me at all times and this report could not have been possible without those. \\
% 
% \noindent Besides my guide, I would like to thank Dr.Vijay Patankar for the wonderful discussions we had, at the Indian Statistical Institute, with fellow interns.  \\
% 
% \noindent My sincere thanks also goes to Dr. Jegannathan.L for his continuous encouragement and motivation to take up this internship. In fact, it was through one of his courses projects that I had got the idea for taking up this topic.  \\ 
% 
% \noindent Additionally, I also thank the library staff at the Institute of Mathematical Sciences (IMSc) for letting me use the library facilities and also the Indian Statistical Institute for letting me use their computers.  
% \newpage
% 

\begin{abstract} 
\thispagestyle{plain}
\pagenumbering{roman}
\setcounter{page}{2}
\addcontentsline{toc}{chapter}{Abstract}

\noindent Prime numbers play a very vital role in modern cryptography and especially the difficulties involved in factoring numbers composed of product of two large prime numbers have been put to use in many modern cryptographic designs. Thus, the problem of distinguishing prime numbers from the rest is vital and therefore there is a need to have efficient primality testing algorithms. \newline 
Although there had been many probabilistic algorithms for primality testing, there wasn't a deterministic polynomial time algorithm until 2002 when Agrawal, Kayal and Saxena came with an algorithm, popularly known as the AKS algorithm, which could test whether a given number is prime or composite in polynomial time. \\ \\ 
This project is an attempt at understanding the ingenious idea behind this algorithm and the underlying principles of mathematics that is required to study it. In fact, through out this project, one of the major objectives has been to make it as much self contained as possible. Finally, the project provides an implementation of the algorithm using Software for Algebra and Geometry Experimentation (SAGE) and arrives at conclusions on how practical or otherwise it is.    

\end{abstract}

\tableofcontents
\setcounter{page}{3}
\addcontentsline{toc}{chapter}{Contents}

\listoftables
\addcontentsline{toc}{chapter}{List of Tables}

\rmfamily\chapter{Introduction}

\pagenumbering{arabic}
\rmfamily  \normalsize Prime numbers have been studied for a long time because of their fundamental importance in mathematics in general and number theory in particular. Of all the properties of prime numbers that have been widely looked into is an age-old problem, perhaps one of the very few well understood problems in pure mathematics, which is to determine whether a given number is prime or composite.  
\\ \\ 
This problem referred to as the Primality Testing problem is of special interest because of a very vital role it plays in cryptography. The difficulty involved in factoring large numbers composed of a product of two primes, as used in the Rivest-Shamir-Adleman (RSA) crypto system, is one such example where the prime numbers are of at most importance and here, efficient primality testing algorithms need to be in place to ensure that the key used in the RSA encryption system indeed uses only prime numbers.
\\ \\
Primality testing has a long history. One of the very first attempts at the problem was the Sieve of Eratosthenes which finds all the prime numbers until a required $n$. The idea, though simple was ingenious and it only relies on a basic understanding of numbers. It merely involves listing down all the numbers from $2$ to $n$ and then beginning with the first number eliminating its factors. At the end of the exercise the numbers remaining unmarked are the prime numbers.\newline Another way to test whether a given number is prime or composite was formulated; this test was based on the fact that for any given number $n$, if $p$ is a factor of $n$ greater than $\sqrt{n}$ then there exists a factor $k$ such that  $ k < \sqrt{n}$. This fundamental property of numbers led to an algorithm which basically dealt with testing if there exists any number $p \leq \sqrt{n}$ such that $p\mid n$\footnote{Refer section 2.1, theorem 2.1.5 for proof}. \\ \\
Although both the above tests where simple, none of them where feasible since the complexity involved in determining whether a given number $n$ was prime or composite was \textbf{$ \Omega(n)$.}   
\\
An almost 'efficient' test was found with the Fermat's little theorem which states that: Given a number prime $p$ and an $a$ such that $(p,a)= 1$, then $a^{p-1}=1$ (mod $p$). Although the above test could be efficiently computed, it could not be used as such since there were numbers called the Carmichael numbers which satisfied the Fermat's property for all $a$'s.\\ \\
\noindent Even though the direct application of the Fermat's little theorem was discarded, it lay the foundation for many different algorithms which were primarily based on this important theorem. Several randomized algorithms were formulated like the Miller-Rabin Test and the Solvay-Strassen Test but none of them, although they were very efficient, could fit into the category of an 'unconditional deterministic polynomial-time algorithm'.
\\ \\
In August 2002, a major breakthrough was achieved by Agrawal, Kayal and Saxena who, together, proposed an algorithm (popularly known as the AKS algorithm) in their paper "`PRIMES is in P". The algorithm which is based on a slight modification of the Fermat's Little Theorem was the first deterministic polynomial-time algorithm for primality testing.\newline Because of the importance of prime numbers especially in cryptography, it is worth investigating and understanding the algorithm. This project is an attempt in this direction i.e. to study the AKS algorithm, understand the basic principle(s) involved and to implement the same in Software for Algebra and Geometry Experimentation (SAGE).
\newpage

\chapter{Some Prerequisites}

In this chapter we present an overview of the Algebra and Number Theory that will be required in the course of reading this report. The main objective here is to make the report as much self contained as possible. The reader may refer to [1], [2] and [3] for a structured and more formal treatment of the topics. 

\section{Number Theory essentials} This section introduces some definitions, terminology and certain proofs, in Number Theory, which will be useful during course of the reading this report. \\ \\ 
\underline{\textbf{Prime and Composite numbers}} \\ \\ \normalsize
\textbf{Definition 2.1.1}: An integer $p > 1$ is called a \textbf{Prime} number if $p$ has no factors other than $1$ and $p$ itself. \\ \\
\textbf{Definition 2.1.2}: Any number $n$ which is not prime is called \textbf{composite}. \\ \\
\noindent\underline{\textbf{Greatest Common Divisor (GCD)}} \\ \\ 
\textbf{Definition 2.1.3}: For any two integers $a,b$, the greatest common divisor of $a,b$ denoted by $(a,b)$ is the an integer $d$ such that $d\mid a, d \mid b$ and all other common divisors of $a,b$ divide $d$. \\

\noindent With the above definition (2.1.3) of GCD and definition (2.1.1), we say that two numbers $a,b$ are co-prime to each other if the GCD of $a,b$ is equal to 1 or in other words $a,b$ is said to co-prime to each other if their only common divisor is $1$  i.e. $(a,b)=1$. \\ \\

\noindent \underline{\textbf{Congruences}} \\ \\ \normalsize
\textbf{Definition 2.1.4}: For any positive integer $n$ and for any $a,b \in \mathbb{Z}$, we say that $a$ is congruent to $b$ modulo $n$, denoted by $a=b$ (mod $n$), if $n \, \mid \, a-b$. The relation $a=b$ (mod $n$) is called a congruence relation and $n$ is called the modulus of the congruence. \newline If $n \, \nmid \, a-b$ then we write the congruence relation as $a\neq b$ (mod $n$).  \\ \\ 

\noindent \textbf{Theorem 2.1.5}: For any composite number $n$, there exists a prime $p$ such that $p < \sqrt{n}$.
\begin{proof} Since $n$ is a composite number, it can be written of the form $a\cdot b$ for some $a,b \in \mathbb{Z}$. Now, we know that both $a$ and $b$ cannot be greater than $\sqrt{n}$. Therefore let us assume that one of them is greater than $\sqrt{n}$ i.e let $b > \sqrt{n}$. Hence we have, 
\begin{center}
$n = a \, b$ \newline \hspace*{-47mm} since $b > \sqrt{n}$, we have $ n = a \, b > a \, \sqrt{n}$ $\Rightarrow \sqrt{n} > a$. \\ \end{center}
\noindent \\ Hence from above we have $ a < \sqrt{n} $ or in other words we have proved that for any factor $b > \sqrt{n}$ we have a corresponding factor $a$ such that $ a < \sqrt{n} $. Now, $a$ can either be prime or composite. If $a$ is composite then let $p$ be a prime factor of $a$ and since $ a < \sqrt{n} $ this implies $ p < \sqrt{n} $. On the other hand if $a$ is prime then $a=p$. \\    \end{proof} 

\noindent \underline{\textbf{Residue Classes}} \\ \\ \noindent \textbf{Definition 2.1.6}: Consider the equivalence relation $ \cdot = \cdot$ (mod $n$). Now, this relation is an equivalence relation on the set $\mathbb{Z}$ and it partitions $\mathbb{Z}$ into equivalence classes. For any $a \in \mathbb{Z}$, the equivalence class containing $n$ is denoted by $[a]_{n}$ and an element $z$ is said to belong to $[a]_{n}$ if  $ z = a $ (mod $n$). So in effect $[a]_{n} = a + m \,\mathbb{Z} $. All these equivalence classes are called \textbf{residual classes} modulo $n$.\\
Also, it is easy to check that for a positive integer $n$, there are precisely $n$ distinct residue classes modulo n, namely, $[a]_n$ for $a = 0, . . . , n - 1$. The set of residue classes modulo $n$ is denoted by $\mathbb{Z}_{n}$.

\noindent \\  \textbf{Definition 2.1.7}: The element $[0]_n \in \mathbb{Z}_{n}$ is called the additive identity since for any $\alpha = [a]_{n} \in \mathbb{Z}$, $ \alpha + [0]_n = [a]_{n} + [0]_n = [a]_n = \alpha$  

\noindent \\ \textbf{Definition 2.1.8}: The element $[1]_n \in \mathbb{Z}_{n}$ is called the multiplicative identity identity since for any $\alpha = [a]_{n} \in \mathbb{Z}$, $ \alpha \cdot [1]_n = [a]_{n} \cdot [1]_n = [a]_n = \alpha$  

\noindent \\ \textbf{Definition 2.1.9}: For any $\alpha, \beta \in \mathbb{Z}_n$, $\beta$ is called as the multiplicative inverse of $\alpha$ if $\alpha \, \beta = [1]_n$ or in other words $\alpha \, \beta = 1$ (mod $n$). \\Also, it is easy to see that a multiplicative inverse for an element $a$ exists if and only if$(a,n)=1$. \\ \\

\noindent \fbox{%
        \parbox{1\linewidth}{\noindent \textbf{Notation}: $\mathbb{Z}_{n}^* $ denotes the set of all elements of $\mathbb{Z}_{n} $ which have a multiplicative inverse } } \\ \\ \\
        
\noindent \underline{\textbf{Euler's Phi function}} \\ \\ 
\noindent \textbf{Definition 2.1.10}: Euler's phi function also known as the Euler's totinet function is defined as the number of elements in the set $\mathbb{Z}_{n}^* $. Alternatively, it can also be defined as the number of elements which are co-prime to a given number $n$. The totient function of a number $n$ is denoted by $\phi(n)$. \\

\noindent For any number $n = p_{1}^{e_1} \,p_{2}^{e_2} \, . . . . . p_{r}^{e_r}$, where $p_1, p_2, . . . ,p_r$ are prime,
\begin{center}
$\phi(n) = n\cdot [(1-\frac{1}{p_1}) \, (1-\frac{1}{p_2}) . . . (1-\frac{1}{p_r})]$. 
\end{center}
\noindent Proof of the above claim can be found in [1].  \\  

\noindent \textbf{Definition 2.1.11}: The \textbf{multiplicative order} of a number $a$ modulo $n$ is defined as the smallest integer $k$ such that $ a^k = 1$ (mod $n$ ). \\ \\

\noindent \underline{\textbf{Euler's theorem and Fermat's little theorem}}\footnote{ The proofs of both the Euler's and the Fermat's little theorem can be found in [1], [2]} \\

\noindent \textbf{Theorem 2.1.12: Euler�s Theorem}: Let $n$ be any positive integer and let $a \in  \mathbb{Z}_{n}^*$ then, $a^{\phi(n)} = 1$ (mod $n$ ).In other words, the multiplicative order of $a$ mod $n$ divides $\phi(n)$( i.e. $O_n (a)) \mid \phi(n)$). \\ 

\noindent A consequence of the Euler's theorem is the Fermat's theorem little theorem, which states that: \newline \\
\noindent \textbf{Theorem 2.1.13: Fermat's Little Theorem}: Let $p$ be a prime number and let $a \in  \mathbb{Z}_{p} - \{ [0]_p \}$ then,   $a^{p-1} = 1$ (mod $p$). \\ \\The Fermat's theorem is a very important result which is used in many primality testing algorithms.\\

\section{Algebra essentials} In this section we introduce the notion of abelian groups, rings and finite fields and a few important properties of each of them which will be used in the subsequent chapters.\\ \\

\noindent\underline{\textbf{Abelian groups}} \\ \\ 
\noindent \textbf{Definition 2.2.1}: An abelian group is defined as a set $G$ with a binary operation $*$ defined on $G$ such that: 
\renewcommand{\theenumi}{\roman{enumi}}
\begin{enumerate}
	\item  $*$ is associative i.e. for any $a,b,c \in G$, $a*(b*c) = (a*b)*c$
	\item there exists an element $e \in G$ called the identity element such that for any $a \in G$, $a*e = e*a = a$.
	\item for every $a \in G$ there exists an $a'$ called the inverse of $a$ such that $a*a' = e =a'*a $. 
  \item the binary operation $*$ defined on $G$ is commutative i.e. $a*b = b*a $
\end{enumerate}

\noindent \\ Although an abelian group is a smaller notion compared to the a 'Group' in general, we do not define the group and its properties as such since, all the groups which would be dealt with in this report are abelian groups. As a matter of fact, a group can be simply defined as above with an exception that the property (iv) given above is not true for a group. In fact, this is the property which distinguishes between a Group and an Abelian Group.   \\ \\

\noindent \fbox{%
        \parbox{1\linewidth}{\textbf{Note:} From the above definition it is easy to see that both $ \mathbb{Z}$ and $\mathbb{Z}_{n}$ form an abelian group under the binary operation addition while, $\mathbb{Z}_{n}^*$ forms an abelian group under multiplication.}} \\ \\ \\
        
\noindent \textbf{Definition 2.2.2} Let $G$ be an abelian group, and let $H$ be a non-empty subset of $G$ such that
\renewcommand{\theenumi}{\roman{enumi}}
\begin{enumerate}
\item for $a,b \in H$, $a + b \in H$. 
\item for $ a \in H$, $-a \in H$.
\end{enumerate}
Then $H$ is called a \textbf{subgroup} of $G$. \\ \\

\noindent \textbf{Definition 2.2.3}: The order of a group $G$ is defined as the number of elements in it. It can be finite or infinite and is denoted by $|G|$. \newline \\

\noindent\underline{\textbf{Cosets and quotient groups}} \\ \\ 
\noindent Now, just like we defined in the case of integers in section $2.1$, here we generalize the notion of congruences to abelian groups. \\ \\
\noindent \textbf{Definition 2.2.4}: Let G be an abelian group, and let H be a subgroup of G. For $a, b \in G$, we write $a = b$ (mod $H$) if $a-b \in H$. From this generalization it is easy to see that the relation $\cdot = \cdot $ (mod $H$) is an equivalence relation on the set $G$. And since this is an equivalence relation, it partitions $G$ into equivalence classes. For any element $a \in G$ we can see that the equivalence class containing $a$ is the set $[a]_{H} = a + H = \{ a + h \, | \, h \in H \,\}$ and these equivalence classes are called \textbf{cosets }of $H$ in $G$.\\

\noindent From the above definition we can see that the number of elements i.e. the cardinality of each of the cosets of $H$ in $G$ is the same and this is same as the number of elements in $H$ or put in other words the cardinality of each of the cosets $=$ order ($H$) = $| H |$. \newline Another observation that can be made is that, since the equivalence relation partitions $G$ into equivalence classes we can say that the cardinality of each of cosets $=$ order of $H$ ($|H|$) will divide order of G ($|G|$) because we need to have $ |H| * m = |G|$ where $m$ is the total number of cosets of $h$ in $G$. This observation precisely translates as the Lagrange's theorem. \\ 

\noindent \textbf{Theorem 2.2.5: Lagrange's Theorem}: Let $G$ be an abelian group and let $H$ be a subgroup of $G$. Then the order of $H$ divides the order of $G$.
\begin{proof} Refer the paragraph above for an explanation. \newline \end{proof} 

\noindent \textbf{Definition 2.2.6}: Consider the set of all cosets of $H$ in $G$. This set is called the \textbf{quotient group} of $G$ modulo $H$ and is denoted as $G/H$. \\ \\

\noindent\underline{\textbf{Rings}} \\ \\ 
\noindent \textbf{Definition 2.2.7}: A commutative ring with unity is a set $R$ together with two operations called addition and multiplication defined on $R$ such that 
\renewcommand{\theenumi}{\roman{enumi}}
\begin{enumerate}
\item the set $R$ forms an abelian group under addition with the additive identity as $0_R$.
\item multiplication is associative on the set $R$ such that for $a,b,c \in R$, $a\,(b \, c) = (a\, b)\, c$.
\item the operation of multiplication distributes over addition i.e. $a \,(b+c) = a\,b + a\,c $.
\item there exists a multiplicative identity $1_{R}$.
\item multiplication is commutative i.e for $a,b \in R$, $a\, b = b\, a$. 
\end{enumerate}
 \noindent \\ Again as in the case of groups, there does exist a more general notion of Rings. But the reason we give the definition of commutative rings with unity is because we would be using only those here. \\ 

\noindent \noindent \textbf{Definition 2.2.8}: Let $R$ be a ring and let $a \in R$ then, we call $a$ as a \textbf{unit} if there exists a $b \in R$ such that $ a \, b = 1_R$. Here $b$ is called the multiplicative inverse of $a$ and vice-versa. \\ \\

\noindent\underline{\textbf{Fields}} \\ \\
\noindent \textbf{Definition 2.2.9}: Let $R$ be a ring. If all the non -zero elements in the ring $R$ have multiplicative inverses then the ring $R$ is called a Field. \\ 
 
\noindent\underline{\textbf{Polynomial rings}} \\ \\
\noindent \textbf{Definition 2.2.10}: Let $R$ be a ring, then we can form the ring of polynomials denoted by $R[X]$ which is the set of all polynomials of the form $a_0 + a_1 X + a_2 X^2 + . . . . + a_n X^n$ where $a_0, a_1, . . . , a_n \in R$. Here the $X$ is not a variable but is indeterminate. \newline Put in other words we say a polynomial $f(X) \in R[X]$ if the coefficients of $f(X) \in R$. \\ 

\noindent\underline{\textbf{Cyclotomic Polynomials}} \\ \\
\noindent \textbf{Definition 2.2.11}: Consider the polynomial $x^n -1$. The factors of this polynomial are $e(\frac{k}{n})$ where $e(t) = e^{2\pi it}$. Therefore this polynomial can be written as: 
\begin{center}
\noindent \\ $x^n -1 = \prod\limits_{k=1}^{n} {\left[ \, x - e\left(\frac{k}{n}\right)\, \right] }$ 
\end{center}
\noindent \\ Now, consider the set $C_{n} = \{ k : 1 \leq k \leq n, (k,n)=1 \}$. We define the cyclotomic polynomial $\Phi_{n}$ for all $n \geq 1$ as:
\begin{center} $\Phi_{n} = \prod\limits_{k \in C_{n}} {\left[ \, x - e\left(\frac{k}{n}\right)\, \right] }$  \end{center}
It is easy to see that the cyclotomic polynomial $\Phi_{n}$ is a monic polynomial (i.e. polynomial with leading coefficient $1$) and has a degree of $\phi(n)$. \\ \\

\section{Additional Theorems} 

\noindent \\ \textbf{Theorem 2.3.1}: Given $n$ is prime, prove that $ n \choose i$ $= 0$ (mod $n$)
\begin{proof} We know that $ n \choose i$ $= \dfrac{ (n-i-1) \,\cdots(n-1)\,(n)}{i!} = \dfrac{ (n-i-1) \,...(n-1)\,(n)}{1\,2\, \cdots i}$. \newline Now, if we assume that the given statement is wrong then, the $n$ in the numerator should be divisible by one of $1$ or $2$ or $\cdots i$. But then, if that is true then $n$ cannot be prime. Hence our assumption is false and therefore we can conclude that $ n \choose i$ $= 0$ (mod $n$) when $n$ is prime. \\ \end{proof}

\noindent \textbf{Theorem 2.3.2}: $ 2n+1 \choose n$ $> 2^{n+1}$, for all $n \geq 2$ 
\begin{proof} For $n=2$, we can see that $5 \choose 2$ $ > 2^{3}$.
Now, we will prove this theorem by induction. Let this be true for some value say, $k$, $k > 2$. So, now we have $ 2k+1 \choose k$ $> 2^{k+1}$. Now, we have to prove that this is true for $k+1$. This implies that we have to show that $ 2k+3 \choose k+1$ $> 2^{k+2}$. This implies $ \dfrac{(2k+3)!}{(k+1)! \, (k+2)!} > 2^{k+2}$. \\

\noindent But, from our assumption that it is true for $n=k$, we have: $ 2k+1 \choose k$ $> 2^{k+1}$. Therefore the it remains to prove that $ \dfrac{(2k+2) \, (2K + 3)}{(k+1) \, (k+2)} > 2$. This is true for all $k > 1$. Therefore our assumption that it is true for $n=k$ is true and therefore the theorem is proved by the principle of mathematical induction. \\  \end{proof}

\chapter{The AKS algorithm} 
The objective here is to describe the Primality Testing Algorithm which, when given an input $n$ outputs whether the given input $n$ is prime or composite. Since the aim is to find a polynomial-time algorithm, the final algorithm should output a result (of whether $n$ is prime or composite) in $ O(\log ^{k}{n})$ where $\log {n}$ refers to $\log_{2} {n}$.  
\section{Basic Idea}
The AKS algorithm is primarily based on a polynomial generalization of the Fermat's Little Theorem which states that:\newline \newline
\textbf{Lemma 3.1.}: Let $a \in \mathbb{Z}_{n}^* $ and let $ n \in \mathbb{N}$. Then $n$ is prime if and only if \newline
\begin{center}
$(X+a)^n = X^n + a$ (mod $n$) \\  
\end{center} 
\begin{proof} Given $n$ is prime. Then since $n \choose i$ $= 0$ (mod $n$)\footnote{refer section 2.3, theorem 2.3.1} for all $i$ therefore all the coefficients in the equation become zero. Hence the theorem is satisfied.\\
%==================================================================================================================
% Proof of that CLAIM add. Correct the appropriate section number in the footnote

Now consider that $(X+a)^n = X^n + a$ (mod $n$), therefore it needs to be proved that any $n$ satisfying the above equation should be prime. Let's assume the contrary. Let $n$ be a composite number. Now, consider a $k$ such that $k\mid n$ and also co-prime to $a$. Here the coefficient of $X^{n-q}$ which is $n \choose k$ $\neq 0$ (mod $n$). Therefore, our assumption that $n$ is composite is false. Hence $n$ should be a prime number. \\ \end{proof}
Although the above lemma could be used as a primality test, it is not done since the process of determining whether the given input $n$ is prime or composite would involve computing all the $n$ coefficients, in the worst case. Therefore the time complexity here would be \textbf{$ \Omega(n)$} which is not polynomial-time.\\ \\ 
Therefore, in order to reduce the complexity involved we take the above lemma but divide both sides of the equation by $X^r-1$ for a chosen $r$ so that the number of computations needed to be performed is less. Hence, the main objective now is to choose an appropriately small $r$ and test if the equation  
\begin{center}
$(X+a)^n = X^n + a$ (mod $X^r-1$,$n$) \\  
\end{center}
\noindent \\ is satisfied for sufficient number of $a$'s.
\section{The Algorithm}
Below is the algorithm proposed by Agrawal, Kayal and Saxena for primality testing:\\

\begin{center}
\begin{enumerate}
	\item If $n=a^b$ for $a \in \mathbb{N}, b>1 $ output COMPOSITE
	\item Find the smallest $r$ such that $O_{r} (n) > \log^2 n$
	\item If $1< (a,n) < n $ for some $a\le r$ then output COMPOSITE
	\item If $n \le r$ output PRIME
	\item For $a=1$ to $\lfloor \sqrt{\phi (n)} \, \, \log n \rfloor $ do \\
              \hspace*{30pt} if $(x+a)^n \neq x^n + a$ mod $(x^r -1, n)$ \\
                                                         \hspace*{75pt} output COMPOSITE  
  \item output PRIME
\end{enumerate}
\end{center} \vspace{6mm}

\noindent A closer look at the algorithm reveals that a slight modification can be done in the step $2$, step $3$ and step $4$ .\\ \\ The following are the modifications: \newline
  \indent $2.$ Find the smallest $r$ such that either \newline
  \hspace*{30mm} $(n,r) > 1$ \newline
  \hspace*{35mm} or \newline
  \hspace*{20mm} $(n,r)=1$ and $O_{r} (n) > \log^2 n$ \\ \\
  \indent $3.$ if $r\geq n$ output PRIME \\ \\
  \indent $4.$ if $(n,r) > 1$ output COMPOSITE \\ \\ \\
  \noindent The advantage of doing the above mentioned modifications is that it reduces the computational time involved in finding an appropriate $r$, although the computational complexity of the whole algorithm remains unchanged. In fact, if $(n,r)=1$ then it suffices to check for an $r$ in the range\footnote{refer section 3.3, Note 2} $(\log^2 {n}, \log^5 {n})$ such that $O_{r} (n) > \log^2 n$.

\section{Proof of Correctness}
In this section we present the proof of the algorithm through a serious of theorems and lemmas\footnote{proofs presented here \textbf{\textit{may not be the same}} as in the original paper. Refer [4] for the original proofs as given by Agrawal, Kayal and Saxena.}. \newline In order to prove that the algorithm is indeed correct we need to prove only one theorem.\newline \\ 

\noindent \textbf{Theorem: The algorithm returns PRIME if and only if the \\ \hspace*{21mm} input $n$ is prime.} 
\begin{proof}In order to prove the above theorem we present the following lemmas.\\ \end{proof} 
\noindent \newline \textbf{Lemma 3.3.1}: Given $n$ is prime, the algorithm will return PRIME.
\begin{proof} If $n$ is prime then the return statements in step $1$ and step $3$ will not be evoked. Now, in the step $5$ of the algorithm the equation would never turn out to be false since by lemma 3.1 (proved earlier) it is satisfied for all prime $n$. Therefore it will return PRIME. \\ \end{proof}

\noindent Therefore having proved the above Lemma, the only thing that remains to be shown is the converse i.e. - if it returns PRIME then $n$ is a prime number. To prove this we make use of the following lemmas.\\ Our first objective is to show that there exists an $r$ in a definite range which satisfies the condition that $O_{r} (n) > \log^2 n$. This is a crucial step since both the important steps $2$ and $5$ (important in terms of taking the maximum computational time) depend on this value of $r$. \\ \\ 

\noindent \textbf{Lemma 3.3.2}: There exists an $r \in (0,\log^5 {n}]$ such that $O_{r} (n) > \log^2 n$, $n>2$.
\begin{proof} For n=2, we can see that the smallest such $r$ which satisfies $O_{r} (n) > \log^2 n$ is $ r=3$. Therefore we need to see only for $n>2$. \newline 
Now, let us assume that there exists no such $r$ which satisfies $O_{r} (n) > \log^2 n$ in the given interval. That is assume that for all $r\in (0,\log^5 {n}] $, $O_{r} (n) \le \log^2 n$ \\
\noindent Next, consider $ \prod\limits_{p \le N} {p} $ i.e consider the product of all prime $p$, such that $p\le N $ : $N= \log^5 n$ \\ 

\noindent Now from the Prime Number theorem, we know that $e^N \le  \prod\limits_{p \le N} {p} $. \newline Also since we have assumed that for all $r\in (0,\log^5 {n}] $, $O_{r} (n) \le \log^2 n$,therefore $  \prod\limits_{p \le N} {p}  \, \mid \,  \prod\limits_{i=1}^{\log^2 n} (n^{i} -1)$. \\ \\ 

\noindent Therefore, we now have $e^N \, \le  \prod\limits_{p \le N} {p} \, \, \le \, \, \prod\limits_{i=1}^{\log^2 n} (n^{i} -1)$. \\ \\ 

\noindent But, $ \prod\limits_{i=1}^{\log^2 n} (n^{i} -1) <  \prod\limits_{i=1}^{\log^2 n} (n^{i}) <  n^{\frac{(\log ^2 n)\cdot (\log ^2 n +1)}{2}} < n^{\log^4 n}$ \newline \\

\noindent Also, we know that $n^{\log^4 n} \le 2^{\log^5 n}$ \\ 

\noindent Therefore, combining all the above results we have \newline 

\noindent $e^N \le  \prod\limits_{p \le N} {p}  \le \prod_{1}^{\log^2 n} (n^{i} -1) < n^{\log^4 n} \le 2^{N} $  \\ \\ which is a contradiction! This arises because of the fact that we assumed there exists no $r\in (0,\log^5 {n}]$ such that $O_{r} (n) > \log^2 n$. Hence we can conclude that there indeed exists at least one $r\in (0,\log^5 {n}]$ such that $O_{r} (n) > \log^2 n$. \newline \newline \end{proof} 
\noindent \newline
\fbox{%
        \parbox{1\linewidth}{\noindent \textbf{Note 1}: A closer examination using Wolfram Alpha reveals that the bound on $r$ can be relaxed to $r\le \log^{4.7} {n} $. \\ \\ \textbf{Note 2}: If $(r,n)=1$ then we need to check for an $r$ only in the range $(\log^2 n, \log^5 n] $ since by Euler's Theorem we have $O_r {(n)} \, | \, \phi(r)$ and therefore $ r \, \geq \, O_r {(n)} \, \geq \, \phi(r) \, > \, \log^2{n}$ }
}

\noindent \\ \\ \\ Now, we know that $ O_r {(n)} > 1$. Therefore there must exist a prime $p$ which divides $n$ such that $ O_r {(p)} > 1$. Let us assume that $p > r$ for, if it is otherwise then step $3$ or step $4$ of the algorithm will find that $n$ is composite. Since, we assume that $p>r$ therefore, we have $(n,r)=1 \, , \, (p,r)=1$ i.e. $p, n \in \mathbb{Z}_{r}^* $. \\ \\  

\noindent So, having made the above observation we now go to step $5$ of the algorithm where $l = \lfloor \sqrt{\phi (n)} \, \, \log n \rfloor $ equations need to be verified. Let us assume that it does not output COMPOSITE in this step, so we have: 
\begin{center}
 $(X+a)^n = X^n + a$ (mod $X^r -1, n$)
\end{center}
for all $a$, $0 \leq a \leq l$.\\ \\ 

\noindent Now, since $p \mid n$ we have:  
\begin{center}
 $(X+a)^n = X^n + a$ (mod $X^r -1, p$)  
\end{center}
for all $a$, $0 \leq a \leq l$. Also by Lemma 3.1 we know that for any prime $p$ 
\begin{center}
 $(X+a)^p = X^p + a$ (mod $X^r -1, p$)
\end{center}
for all $a$, $0 \leq a \leq l$. So from the above two equations we have: 
\begin{center}
 $(X+a)^{\frac{n}{p}} = X^{\frac{n}{p}} + a$ (mod $X^r -1, p$)
\end{center}
for all $a$, $0 \leq a \leq l$. \\ \\ 

\noindent Here, we define an idea called 'Introspective Numbers' as they call it.\\ \\ 
\fbox{%
        \parbox{1\linewidth}{\noindent \textbf{Definition}: A number $m$ is called \textbf{introspective} if $(X+a)^m = X^m + a$ (mod $X^r -1, p$) }
}
    
\noindent \\ So, from above we can see that both $p$ and $\frac{n}{p}$ are introspective. In fact, this property can be generalized as a theorem. \\ \\ 

\noindent \textbf{Theorem 3.3.3:} The set of introspective numbers is closed under multiplication 
\begin{proof} Let $m_1 , m_2$ be two introspective numbers introspective to a polynomial $f(X)$. So we need to prove that $m_1 \cdot m_2$ is also introspective. Since $m_1$ is introspective, we have: 
\begin{center}
 $[f(X)]^{m_1 \cdot m_2} = [f(X^{m_1})]^{m_2} + a$ (mod $X^r -1, p$)
\end{center}
\noindent   We also know that $m_2$ is introspective. Therefore, we replace $X$ by $X^{m_1}$ to have: 
\begin{center}
 $[f(X^{m_1})]^{m_2} = f(X^{m_1 \cdot m_2}) + a$ (mod $X^{r\cdot m_1} -1, p$) \newline $= f(X^{m_1 \cdot m_2}) + a$ (mod $X^{r} -1, p$)
 \end{center}

\noindent Therefore from above we have: 
\begin{center}
 $[f(X)]^{m_1 \cdot m_2}= f(X^{m_1 \cdot m_2}) + a$ (mod $X^{r} -1, p$) 
 \end{center} 
\noindent  \newline 
\end{proof}

\noindent \newline Similarly we can have one more theorem which states that: 
\\ \\ 
\noindent \textbf{Theorem 3.3.4:} The set of polynomials for which a number $m$ is introspective is closed under multiplication 
\begin{proof} Let $f(X), g(X)$ be the two polynomials introspective with respect to $m$. Therefore we have: 
\begin{center}
 $[f(X) \cdot g(X)]^{m}= [f(X)]^{m} \cdot [g(X)]^m$ (mod $X^{r} -1, p$) \newline \hspace*{9mm} $ = [f(X^m)] \cdot [g(X^m)]$ (mod $X^{r} -1, p)$
\end{center} 
\end{proof}

\noindent  Having proved the above two theorems, we now move to the what can be called as the main part of the proof. This involves constructing of two groups $G_1$ and $G_2$.\\ \\ 

\noindent Let $ I = \{ (\frac{n}{p})^{i} \, \, (p)^{j} \, \, | \, \, i,j \geq 0\}$ denote the set of all the numbers which are introspective to the polynomials in the set, say $ P = \{ \prod\limits_{a=0}^{l} (X+a)^{e_a} \, | \, e_a \geq 0 \}$ (This follows from the two theorems above).
\\ \\ 

\noindent Now, construct a group $G_1$ of the residues of $I$ modulo $r$. This is a subgroup of  $\mathbb{Z}_{r}^*$ since $(n,r) =1$ and $(p,r)=1$. Let $|G_1| = t$. Since the $O_r (n) > \log ^2 {n} $ therefore, $ t > \log ^2 {n} $. \\

\noindent Before, we construct the second group we state a important theorem (without proof)\footnote{refer [3] for the proof}.\\ \\ 
\noindent \textbf{Theorem 3.3.5:} Consider the $r^{th}$ cyclotomic polynomial\footnote{refer section 2.2. definition 2.2.11} $Q_r (X)$ over the finite field \footnote{refer section 2.2} $\mathbb{F}_{p}$. $Q_r (X)$ divides the polynomial $X^r -1 $ into irreducible factors of degree $> O_r (p)$. \\ \\
\noindent As a consequence of the above theorem if $h(X)$ is one such irreducible factor then the degree of $h(x) > 1$ since the $ O_r (p) >1$.\\ \\ 
\noindent Now, to construct the second group $G_2$ consider the residues of all the polynomials in $P$ modulo $p$ and $h(X)$. This group $G_2$ will be generated by $ X , X+1, X+2,\cdots , X+l$ in the field  $F = \mathbb{F}_{p}[X] / h(X) $. \\

\noindent After having constructed the two groups we have the following lemmas. \\ \\ 

\noindent \textbf{Lemma 3.3.6:} $ |\,G_2 \,|$  $\geq$ $ t+l \choose t-1 $ 
\begin{proof} Consider two polynomials $f(X)$ and $g(X)$ both in $P$ of degree $< t$. First, we show that both $f(X)$ and $g(X)$ map differently in $G_2$. For that, let us assume $f(X) = g(X)$ in $F$ and let $m \in I$. Since $m$ is introspective to both $f(X)$ and $ g(X)$, we have: 
\begin{center}
 $f(X^m) = g(X^m)$ in $F$
\end{center}
Hence replacing $Y=X^m$ we see that $Q(Y) = f(Y) - g(Y)$ has roots of the form $X^m$ and this is true for all $m \in G_1$. Therefore we know that the polynomial $Q(Y)$ has $| G_1 | = t$ distinct roots in $F$. But then, both $f(Y)$ and $g(Y)$ are of degree $< t$, which implies that it is impossible for $Q(Y)$ to have $t$  distinct roots. This is contradiction and has arised since we assumed that $f(X) = g(X)$ in $F$. Therefore, $f(X) \neq g(X)$  in $F$. \\ 

\noindent Now, $ l = \lfloor \sqrt{\phi (n)} \, \, \log n \rfloor  < \sqrt{r} \, \log n < r$ (since we have assumed that $(n,r)=1$)\footnote{refer section 3.3 Note $2$} and we also know that $ p > r$. Therefore, the elements $ X , X+1, X+2,\cdots , X+l$ are all distinct in $\mathbb{F}_{p}$. Also, since the degree of $h(X) > 1$ therefore $X+a \neq 0$ in $F$ for any $ a \in [0,l]$. This shows that there are at least $ l+1$ polynomials of degree one in $G_2$. Hence, the number of polynomials with degree $ < t$ is: \newline 
$=$ no: of polynomials of the form $ (X)^{e_0} \cdot (X+1)^{e_1} \cdots (X+l)^{e_l}$ with degree \hspace*{3mm} $ < t $ \newline 
$=$ no: of solutions of : $ e_0 + e_1 + e_2 + \cdots + e_l < t$ \newline \newline
$=$ $ t+l \choose t-1 $ \\  

\noindent Therefore the number of elements in $ G_2$ has to be greater than $ t+l \choose t-1 $. Or in other words $ |\,G_2 \,|$  $\geq$ $ t+l \choose t-1 $. \\ 
\end{proof}
\noindent With a lower bound on the number of elements in $|G_2|$ we can now show that if $n$ is not a power of a prime number $p$ then we can arrive at an upper bound on the number of elements. This is illustrated in the following lemma: \\ \\ \\
\noindent \textbf{Lemma 3.3.7:} If $n$ is not a power of $p$ then, $ |\,G_2 \,|$  $\leq$ $ n^{\sqrt{t}} $ 
\begin{proof} Consider a subset of the set of introspective numbers $I$ say $I^* = \{ {(\frac{n}{p}})^i\, \,  p^j \, \,| \, \, 0 \leq \, i,j \leq \, \sqrt{t} \, \, \}$. Now, if $n$ is not a power of prime $p$ then the number of distinct elements in $I^* =  (\sqrt{t} + 1)^2 > t $. This implies that when the elements of $I^*$ are taken modulo $r$ at least, two of them will be equivalent since, we know that the number of elements in $ |G_1| = t$. \\ 

\noindent Now, let the two numbers in $I^*$ which are equivalent modulo $r$ be $m_1, m_2 : \, \,m_1 > m_2$. So we have: 
\begin{center}
 $X^{m_1} = X^{m_2}$ (mod $X^{r} -1$)
\end{center}
\noindent \\ If $f(X) \in P$ then, since both $m_1, m_2 \in I $, we have:
\begin{center}
$[f(X)]^{m_1} = f(X^{m_1})$ (mod $X^{r} -1, p$) \newline \hspace*{14mm} $= f(X^{m_2})$ (mod $X^{r} -1, p$) \newline \hspace*{-8mm} $=  [f(X)]^{m_2}$ (mod $X^{r} -1, p$)

\end{center}
\noindent \\ Therefore from above we have:
\begin{center}
$[f(X)]^{m_1} = [f(X)]^{m_2}$ in $F$.
\end{center}
\noindent \\ And this implies that all $f(X) \in G_2$ is a solution to the polynomial $ Q(Y) = Y^{m_1} - Y^{m_2}$. We also know that this polynomial can have a maximum of $m_1$ roots and the maximum value of $m_1$ is $n^{\sqrt{t}}$. But then, all the $f(X) \in G_2$ are roots of $Q(Y)$ which implies that the maximum value of $|\,G_2 \,|$ is $n^{\sqrt{t}}$ or in other words $ |\,G_2 \,|$  $\leq$ $ n^{\sqrt{t}} $ \\ \end{proof}

\noindent \newline Now, we come to the final part of the algorithm where we prove the statement of the theorem we began with in the section $3$. \\ 

\noindent \\ \textbf{Lemma 3.3.8:} If the algorthim returns PRIME then $n$ is prime.
\begin{proof} Given that the algorithm returns PRIME we know that it can do so only in step $4$ and step $6$ of the algorithm. Having already seen the step $4$ case, we now have to see that the if the algorithm  returns PRIME in step $6$ then $n$ is indeed a prime number. \\ \\
Now, from {Lemma 3.3.6:} we know that  $|\,G_2 \,|$  $\geq$ $ t+l \choose t-1 $. We also know that $ t > \sqrt{t}  \, \log {n} $. Also, by Lagrange's theorem\footnote{refer section 2.2, theorem 2.2.5} we know that order of any finite group is divisible by the order of its subgroup. Therefore since $G_1$ is a subgroup of $\mathbb{Z}_{r}^* $, $ t \, \mid \, \phi ({r}) $. Hence we have $ l > \sqrt{t}  \, \log {n} $. Substituting the above results we have: \\ 
\begin{center}
$|\,G_2 \,|$  $\geq$ $ t+l \choose t-1 $ \newline \newline \hspace*{14mm} $\geq$ $2 \lfloor \sqrt{t}  \, \log {n} \rfloor + 1 \choose \lfloor \sqrt{t}  \, \log {n} \rfloor$ \newline \newline \hspace*{13mm}  $\geq 2^{\lfloor \sqrt{t}  \, \log {n} \rfloor + 1}$ \vspace*{2mm} \newline  \hspace*{35mm} {$\{$ since $2n + 1 \choose n$ $> 2^{n + 1 }$ for $n \geq 2 \}$}\footnote{refer section 2.3, theorem 2.3.2} \newline \newline  \hspace*{-35mm} $ \geq n^{\sqrt{t}}$
\end{center}

\noindent \\ Therefore from above we have $ |\,G_2 \,|$  $\geq$ $ n^{\sqrt{t}} $. This implies that $n$ is a power of prime $p$ because otherwise from Lemma 3.3.7 we would have had $ |\,G_2 \,|$  $\leq $ $ n^{\sqrt{t}} $. But then, if $n$ was a prime power of $p$ that would have been detected in the very first step. Therefore, the only possibility is $n = p^1$ which implies that $n$ is a prime number and hence this proves the correctness of the AKS algorithm. \\ \\
\end{proof}   

%====================================================================== C(2n+1, n) > 2^(n+1) 
\chapter{Implementation}
Having proved the correctness of the AKS algorithm in the previous chapter we now move on to the implementation of the algorithm. In this chapter we look at the various modules which are involved in the implementation of the algorithm and finally we present the algorithm which was implemented in SAGE (Software for Algebra and Geometry Experimentation).\\ The implementation was done by defining a function AKS(n) which takes a number n as the input and outputs whether the number is prime or composite. 

\section{Testing for perfect-power} This part of the algorithm is used to check if the given number $n$ is a perfect power i.e here we check if the number $n$ can be written in the form $a^b$.Now, if $n=a^b$ then we know that the maximum value of $b$ is $\log n$ where log here refers to the base 2 (as has been the case throughout). Therefore the problem here reduces to probing whether an $a$ exists such that $a^b =n$ for $b \in [2,\log n]$. The algorithm for the same is given below: \\
\begin{center}
\begin{enumerate}
	\item for $b=2$ to $\lfloor \log n \rfloor$
	\item \hspace*{8mm} $y= \frac{\log n}{b}$
	\item \hspace*{8mm} $a=$pow($2,y$) // performs $2^y$
	\item \hspace*{8mm}  if ($a^b == n$) \newline \hspace*{18mm} output TRUE
	\item Output FALSE 										
\end{enumerate}
\end{center}

\section{Finding an appropriate $r$} Having checked whether the given number $n$ is a perfect power or otherwise we now move on to the step $2$ of the algorithm, which as mentioned earlier is one of the most important step of the AKS algorithm. The objective here is to find an appropriate $r$ such that $O_r{(n)}>\log^2 n$. Now, in the correctness proof of the algorithm we found that such an $r$ can be found in the range $ (0,\log^5 n]$ so, using that we need find an $r$ which satisfies the above condition. To do so, we test all $ r \in [2,\log^5 n] $ to see if there is a $ k \in [1,\log^2 n]$ such that $n^k = 1$ mod($r$). If there is no such $k$ then that particular $r$ is the appropriate $r$ that we want. Given below is the algorithm for finding such an $r$. \\

\begin{center}
\begin{enumerate}
	\item for $r=2$ to $\lfloor \log^5 n \rfloor + 1$
	\item \hspace*{8mm} if for all $k \in [1,\log^2 n]$ \newline \hspace*{18mm} $n^k \neq 1$ (mod $r$) then output $r$ 
	\item else continue till finding such an $r$
										
\end{enumerate}
\end{center}

\section{Checking $l$ equations} Here we move on to the step $5$ of the algorithm which basically checks $l$ equations. Now, one reason for choosing SAGE to implement the algorithm was because SAGE basically has built-in libraries and routines which can form rings modulo n, form polynomial rings and  perform quotienting over polynomial rings. The routines we used here are Integers(), PolynomialRing() and quotient(). Integers() is used to form the ring of integers modulo any n where n will be given as a parameter to it. The PolynomialRing() forms a polynomial ring over the ring say s, which will be given as a parameter to it and the quotient() is used for quotienting over polynomial rings. \\ 

\noindent Now putting together all the pieces we present the SAGE\footnote{ A good place to look at SAGE commands is [8], [9]. } code for the AKS algorithm:
\begin{verbatim}
def AKS(n):
    c=1
    k=0
    for b in range(2,ceil(log(n,2))+1): 
        y=(log(n,2)/b).n()                        
        c=(pow(2,y)).n(30)    
        if (pow(floor(c),b)==n) :
            print 'Composite'
            print 'It is a perfect power. n =', n, '=',floor(c),'^', b          
            return   
      
    m=((log(n,2)))^5
    r=2
            
    while(r<=floor(m)):
            c=0                         
            i=floor(log(n,2))^2         
            for k in range(1,i+1):             
                if ((n^k - 1)%r==0):                   
                   c=c+1	           
           
           
            if(c==0):            
                 break                                      
            r=r+1             
            
    print 'smallest r is',r 
    
    for a in range(1,r):
        if(1<gcd(a,n)<n):
            print 'composite'
            return   
    
    if(r>=n):
        return 'prime'
        
				    l=floor((2*sqrt(euler_phi(r))*log(n,2)+1))
    
    for a in range(1,l):
        s=Integers(n)
        R.<x>=PolynomialRing(s)
        F = R.quotient((x^r)-1)
        q=F((x+a))
        V=F(q^n)
        e=Mod(n,r)
        d=(x^e)+a
        if (V!=d):
        	   return 'composite'
    return 'prime'
    
\end{verbatim}

\chapter{Results}
Having presented the implementation of the AKS algorithm in the previous chapter, we now present the results obtained on running the code. Here, in order to be sure that the implemented algorithm is correct we test the individual components and check if it gives the right results.
\section{Perfect Power Checking} In order to prove that the algorithm given for perfect power checking is true, we present the table (5.1) below which tests this algorithm for certain values. The values are selected so that, it is tested for numbers with different number of digits. In the result column, a $0$ implies that the entered number is not a perfect power while otherwise, it displays the perfect power of the given number such that $n=a^b$. 
\section{Finding the appropriate $r$} The table below (5.2) shows the values of $r$ for different values of $n$ such that the $O_r {(n)} > \log^2 {n}$. The '$r$' column gives the values of the smallest $r$ which satisfies the above condition.  \\
\section{Testing AKS(n)} Finally, after having checked the correctness of perfect-power algorithm and the value of $r$ for some values of $n$, here we see the output of AKS(n) for certain $n$. Table(5.3) below tabulates the results obtained for certain values of $n$ and in this table the result column indicates whether the number $n$ is PRIME or COMPOSITE. \newpage

\begin{flushleft}
\begin{tabular}{| C{.49in} | R{4in}|R{0.7in}|}
\hline                 
\textbf{Digits}         &\textbf{ Number} �������������������������������������������������������� & \textbf{Result�}� \\ \hline
��������1������������ & 5�������������������������������������������������������������� & $0$����� \\ \hline 
��������1������������ & 8�������������������������������������������������������������� & $2^3$��� \\ \hline 
��������5������������ & 33561���������������������������������������������������������� & $0$����� \\ \hline 
��������5������������ & 50653���������������������������������������������������������� & $37^3$�� \\ \hline 
				11						& 74589621369																											& $0$      \\  \hline 
				11						& 62523502209																											&	$250047 ^ 2$   \\  \hline 
				28						&	1956410986640441413344189841																		&	$7921 ^ 7$	\\  \hline 
				28						& 1331410986640441413344189841																		&	$0$		\\  \hline 
				43						& 1166316484553604833910724976125484023676928											& $6672^{11}$    \\  \hline     
				58						& 2478773645164848478975418774582564813617861031462 \vspace{1mm}649237504			& $0$ \\ \hline
				67						& 6537582183815232936509417459731287452239283984923  \vspace{1mm} 147562890500202607 				& $143^{31}$  \\  \hline 
				80						& 5015772548390522734176174889014169418007610109084  \vspace{1mm} 0037071925071763978352502147041	&$493039 ^ {14}$ \\ \hline 
				132						& 1915028638109830236694870661712657837656061763680 \vspace{1mm} 6731780114577653169832206102664839162775687537780  \vspace{1mm} 7727062582859038489434170143851401& $201^{57}$ \\ \hline				
				206           & 1861897383944053881816687033971150716922455264712 \vspace{1mm} 72490860683088476343948454916507097899181028458949  \vspace{1mm} 70988359834132367548553692165243808101362414308102\vspace{1mm} 61316931856112708001397269161932139328756319703935  \vspace{1mm}7670 1952 &$43614208 ^ {27}$  \\  \hline

\end{tabular} \end{flushleft}
\addcontentsline{lot}{chapter}{5.1 Table showing the correctness of perfect power test}
\noindent
\begin{center}
Table 5.1: Table showing the correctness of perfect power test
\end{center}

\small\begin{center}
\begin{tabular}{| R{1.3in} | C{0.5in}|}
\hline                 
\textbf{Number}         &\textbf{r} \\ \hline		
5												& 5 \\ \hline		
41											&41 \\ \hline		
983											&101 \\ \hline		
2909										&149 \\ \hline
65909										&17	 \\ \hline		
489721									&79 \\ \hline		
8895643									&47 \\ \hline		 
36952741								&7	\\ \hline		
1307135101							&941 \\ \hline
45884698721							&179 \\ \hline	
7000000000000037				&1189  \\ \hline	
\end{tabular}
\end{center}
\noindent
\begin{center}
Table 5.2: Table showing the values of $r$ obtained for a given $n$ 
\end{center}
\addcontentsline{lot}{chapter}{5.2 Table showing the values of $r$ obtained for a given $n$}
\noindent 
\begin{center}
\begin{tabular}{| R{3.1in} | C{1in}|}
\hline                 
\textbf{Number}         &\textbf{Result} \\ \hline		
5												& PRIME \\ \hline		
33											&COMPOSITE \\ \hline		
861											& COMPOSITE \\ \hline		
4861										&PRIME \\ \hline
55697									  &PRIME	 \\ \hline		
7741043								  &PRIME \\ \hline		 
1771561									&COMPOSITE \\ \hline
90552556447							&COMPOSITE	\\ \hline		
435465768763						&PRIME\\ \hline
9965468763136528274628451			&PRIME\\ \hline			
43546576876313652827462842409420126531423 &COMPOSITE \\ \hline	\end{tabular}
\end{center}
\noindent\begin{center}
Table 5.3: Table showing the results obtained for AKS(n)
\end{center}

\addcontentsline{lot}{chapter}{5.3 Table showing the results obtained for AKS(n)}

\normalsize

\chapter{Conclusions}
The formulation of a Deterministic Polynomial Time Primality Testing Algorithm has certainly been a huge leap considering the fact that there where none until then and the AKS algorithm has certainly thrown us some light on one of the oldest problems confronting mathematics, which is to test whether a given number is prime or composite. \\ But having said that, it is also to be reminded that the AKS algorithm though a remarkable theoretical result, is still nowhere near to being practical since we have better performing probabilistic algorithms (with very less margin for error). The very fact that the check for whether the number $9965468763136528274628451$ is prime ( in Table 5.3 ) took as much as around 70 minutes shows how inefficient it is. And although, there has been considerable work undertaken to improve the algorithm, most of these have been based on results which haven't been proven yet. \\ \\
Considering the situations where these primality testing algorithms are used, like in cryptography where they have to confront very large numbers, it is an absolute need of the hour to improve it so as to make it more practical and therefore, I certainly believe that there is still a lot of work to be done in this area. \\  

\noindent Although I do feel that this project would have been better off if it had resulted in some contributions towards making the algorithm more practical, I certainly believe that I have been able to meet the modest target I had set (especially because of the insufficient mathematical background I possessed) - which was to basically study and have a good understanding of the deterministic primality testing. In addition, it has also given me an opportunity to understand some basic Number Theory and Abstract Algebra, most of which, were new to me, and appreciate some results which have huge significance in this and many other related areas. Also, an introduction into SAGE programing, I feel, was a definite positive considering how powerful the tool turned out to be, especially during the implementation phase of this project.

\newpage
\noindent \Huge\textbf{References} \addcontentsline{toc}{chapter}{References} \\ \\ \\ \\
\normalsize
\noindent \\ \textbf{Books:} \\ \\
\noindent [1] \hspace{5mm} Victor Shoup. A Computational Introduction to Number Theory \\ \hspace*{11mm} and  Algebra  \\ \\
\noindent [2] \hspace{5mm} G. H. Hardy and E. M. Wrightan. Introduction to the theory of \\  \hspace*{11mm} numbers pp 63 - 72 \\ \\

\noindent [3] \hspace{5mm} Rudolf Lidl and Harald Niederreiter. Introduction to finite fields \\ \hspace*{11mm} and their applications pp 11 - 36\\ \\

\noindent \\ \textbf{Journal Papers:} \\

\noindent [4] \hspace{5mm} M. Agrawal, N. Kayal and N. Saxena, �PRIMES is in P�  \\ \hspace*{11mm} \url{www.cse.iitk.ac.in/users/manindra/algebra/primality_v6.pdf} \\ \\

\noindent [5] \hspace{5mm} Jaikumar Radhakrishnan, Kavitha Telikepalli, V.Vijay. News from \\ \hspace*{11mm} India: Primes is in P \\ \\

\noindent [6] \hspace{5mm} Andrew Granville, It is Easy to Determine Whether a Given \\ \hspace*{11mm} Integer is PRIME \newline \hspace*{10mm}  \url{http://www.dms.umontreal.ca/~andrew/PDF/Bulletin04.pdf}\\ \\

\noindent [7] \hspace{5mm} Cyclotomic Polynomials: Notes by G.J.O. Jameson \newline \hspace*{11mm} \url{http://www.maths.lancs.ac.uk/~jameson/cyp.pdf} \\ \\

\noindent [8] \hspace{5mm} SAGE help and documentation \url{http://sagemath.org/doc/} \\ \\

\noindent [9] \hspace{5mm} Sage for Abstract Algebra: A Supplement to Abstract Algebra,\\  \hspace*{10mm} Theory and Applications by Robert  A.Beezer \\ \hspace*{10mm} \url{http://abstract.ups.edu/download/aata-20111223-sage-4.8.pdf}
\end{document}